\documentclass[prl,twocolumn,nobalancelastpage,superscriptaddress]{revtex4-2}
\usepackage{amsmath,amssymb,graphicx}

\newcommand{\ket}[1]{\vert#1\rangle}
\newcommand{\bra}[1]{\langle#1\vert}

\begin{document}

\title{Efficient description of many-body systems with Matrix Product Density Operators}

\author{Ji\v{r}\'i Guth Jarkovsk\'y}
\affiliation{Max-Planck-Institute of Quantum Optics, Hans-Kopfermann-Str.~1, 85748 Garching, Germany}
\affiliation{Munich Center for Quantum Science and Technology, Schellingstr.~4, 80799 M\"unchen, Germany}
\author{Andr\'as Moln\'ar}
\affiliation{Max-Planck-Institute of Quantum Optics, Hans-Kopfermann-Str.~1, 85748 Garching, Germany}
\affiliation{Munich Center for Quantum Science and Technology, Schellingstr.~4, 80799 M\"unchen, Germany}
\affiliation{\mbox{Instituto de Ciencias Matem\'aticas, Campus Cantoblanco
UAM, C/ Nicol\'as Cabrera, 13-15, 28049 Madrid, Spain}}
\affiliation{\mbox{Dpto.\ An\'alisis Matem\'atico y Matem\'atica Aplicada,
Universidad Complutense de Madrid, 28040 Madrid, Spain}}
\author{Norbert Schuch}
\author{J.~Ignacio Cirac}
\affiliation{Max-Planck-Institute of Quantum Optics, Hans-Kopfermann-Str.~1, 85748 Garching, Germany}
\affiliation{Munich Center for Quantum Science and Technology, Schellingstr.~4, 80799 M\"unchen, Germany}

\begin{abstract}
Matrix Product States form the basis of powerful simulation methods for
ground state problems in one dimension. Their power stems from the fact
that they faithfully approximate states with a low amount of entanglement,
the ``area law''.  In this work, we establish the mixed state analogue of
this result: We show that one-dimensional mixed states with  a low amount
of entanglement, quantified by the entanglement of purification, can be
efficiently approximated by Matrix Product Density Operators (MPDOs). In
combination with results establishing area laws for thermal states, this
helps to put the use of MPDOs in the simulation of thermal
states on a formal footing.

\end{abstract}

\maketitle

\hyphenation{MPDO}
\hyphenation{MPDOs}

Complex interacting quantum many-body systems cannot be understood without
carefully assessing the structure of their quantum correlations, that is,
entanglement.  This is the key insight behind the Density Matrix
Renormalization Group (DMRG) method for the simulation of one-dimensional
(1D) systems~\cite{white:DMRG}, and it has later allowed to explain DMRG
as a variational method over the manifold of Matrix Product States
(MPS)~\cite{ostlund-rommer,verstraete:dmrg-mps}.  This understanding has in turn
triggered generalizations of the method for instance to the simulation of
time evolution, excitation dynamics, thermal states, to the development of
methods for the simulation two- and higher-dimensional systems, gauge
theories, and so
forth~\cite{cirac:tn-review-2009,schollwoeck:review-annphys,schuch:juelich-notes-feb13,orus:tn-review,bridgeman:interpretive-dance}.

The success of MPS-based methods
stems from the fact that MPS faithfully approximate the relevant states,
in particular low-energy states of physical systems.  The reason for this
is deeply rooted in the entanglement structure of those sytems: Ground
states and low-lying excited states of gapped local Hamiltonians obey an
area law for the entanglement entropy -- that is, for any contiguous region
$A$, the entanglement $E(A:A^c)$ between $A$ and its complement $A^c$ is
bounded by the length of the boundary $\partial A$, $E(A:A^c)\le
\mathrm{const}\times |\partial A|$ (in particular, this is a constant in
1D)~\cite{hastings:arealaw,arad:rg-algorithms-and-area-laws}. Even for
critical systems, this behavior is at most logarithmically violated,
$E(A:A^c)\le \mathrm{const}\times \log |A|$ (with $|A|$ the number of
sites in $A$).  This is a very special property, since the vast majority
of quantum many-body states has essentially maximum entanglement, i.e., a
volume law~\cite{hayden:epsilon-nets}.  This demonstrates that it is precisely
the area law scaling of the entanglement which characterizes the
physically relevant ``corner'' of Hilbert space.

In order to succinctly characterize those states, it is thus necessary to
understand the structure of states with limited entanglement, such as an
area law or with an only logarithmic increase.  It has been found that this is
precisely captured by Matrix Product States
(MPS)
\begin{equation}
\ket{\phi_D} = \!\!\! \sum_{i_1,\dots,i_N=1}^d 
\!\!\!A^{[1]}_{i_1}A^{[2]}_{i_2}\cdots
    A^{[N]}_{i_N}\:\ket{i_1,\dots,i_N}
\end{equation}
(here for a chain of $N$ $d$-level spins),
where the $A^{[k]}_{i_k}$ are $D\times D$ matrices, except for
$A^{[1]}_{i_1}$ and $A^{[N]}_{i_N}$ which are $1\times D$ and $D\times 1$
matrices, respectively.  On the one hand, any MPS obeys an area law by
construction. On the other hand, it has been shown that any state which
obeys a suitable area law for the entanglement can be faithfully
approximated by an MPS, that is, with a $D$ which only grows
\emph{polynomially} in the system size $N$ and the desired accuracy (in
contrast to the Hilbert space dimension
$d^N$)~\cite{verstraete:faithfully,schuch:mps-entropies}.  This clarifies
why MPS are well suited to describe ground states and excitations of
quantum many-body systems, allowing for efficient simulations.

For thermal states, or more generally mixed states, the situation
is much less clear. Again, we can ask the same two key questions: First, what is the
structure of entanglement, or more generally correlations, in thermal
states of physical systems -- in particular, is there some analogue to the
area law?  And if yes, second: Given a mixed state $\rho$ which obeys an area
law, what is the structure of $\rho$, and in particular, can it be
well approximated by a Matrix Product Density Operator (MPDO)
\begin{equation}
\label{eq:mpdo}
\sigma_D =\!\!\sum A^{[1]}_{i_1j_1}A^{[2]}_{i_2j_2}\cdots
    A^{[N]}_{i_Nj_N}\ket{i_1,\dots,i_N}\bra{j_1,\dots,j_N}
\end{equation}
(where the sum runs over $i_k,j_k=1,\dots,d$, and the $A^{[k]}_{i_kj_k}$ are
as before)?
The problem is further impeded by the fact that for mixed states there exists
a whole zoo of different entanglement measures which are often not related
in a simple fashion.  As it turns out, the first of those questions has
been addressed previously: It has been shown that
thermal states of local Hamiltonians obey an area law for the mutual
information, which quantifies both quantum and classical correlations in
the system~\cite{wolf:mutual-info-arealaw}.   The second question, however
-- relating entanglement scaling and approximability by MPDOs -- is yet
open, and this is what the present work
deals with.

In this paper,  we show that also for mixed states, a suitable 
family of entanglement area laws -- even with a logarithmic correction --
implies that the state can be efficiently approximated by MPDOs.
Specifically, given a mixed state $\rho$ on a chain of $N$ $d$-level
spins, we prove that if there exist constants $c>0$ and
$0<\lambda<1$ such that the \emph{$\alpha$-R\'enyi entanglement of
purification} $E_{p,\alpha}$ (defined
below)~\cite{terhal:entanglement-of-purification} satisfies
\begin{equation}
\label{eq:eop-arealaw-bound}
E_{p,\alpha}(A:A^c) \le c \log N\mbox{\quad for\ }
    \alpha = \frac{\lambda}{5\log_2 N}\ ,
\end{equation}
then $\rho$ can be efficiently approximated by an MPDO $\sigma_D$: As long
as the bond dimension $D$ scales polynomially, $D=N^\kappa$ for any
$\kappa>\tfrac{2c}{1-\lambda}$, the error
$\varepsilon:=\|\rho-\sigma_D\|_1$ in trace
norm goes to zero super-polynomially in $N$ (i.e., faster than any inverse
polynomial)~\footnote{Note that due to the monotonicity of the
$\alpha$-R\'enyi entropy in $\alpha$, we can also choose any smaller
$\alpha$  in \eqref{eq:eop-arealaw-bound}.}.  By using the trace norm --
which exactly bounds the error in expectation values of arbitrary bounded
observables (with largest eigenvalue $1$) -- we obtain a bound on the
error incurred in arbitrary simulations of physical processes.  This
establishes that MPDOs are precisely the framework needed to faithfully
describe mixed states which obey an entanglement area law of the form
above.

\begin{figure}[b]
\includegraphics[width=8cm]{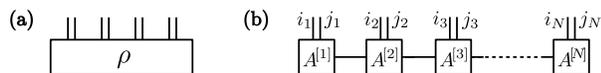}
\caption{\textbf{(a)} Tensor notation of density matrix $\rho$: Each pair
of legs denotes the ket and bra index at one site. \textbf{(b)} Tensor
network for MPDO, Eq.~\eqref{eq:mpdo}. Legs denote indices, connected lines contraction
(summation) of indices, corresponding to the matrix products in
Eq.~\eqref{eq:mpdo}.}
\label{fig:tn}
\end{figure}

The proof will on the one side follow the approach in the pure state
case~\cite{verstraete:faithfully}. On the other side, we need to 
deviate from it at some key steps.  The reason is that we require a good
approximation in trace norm, which -- unlike the $2$-norm -- does not
induce a scalar product, which in turn is essential to build
norm-preserving projections. At the same time, we cannot
bound the $2$-norm instead: The relative bound $\|\rho\|_1 \le\sqrt{\mathcal
D}\|\rho\|_2$, with $\mathcal D=d^{N}$ the dimension of the total space,
is tight (saturated by the maximally mixed state), that is, the trace norm
can be exponentially larger in $N$ than the $2$-norm, breaking efficiency
of the approximation.

We will use the conventional graphical calculus for
MPS/MPDOs~\cite{cirac:tn-review-2009,schollwoeck:review-annphys,schuch:juelich-notes-feb13,orus:tn-review,bridgeman:interpretive-dance},
where a (mixed) many-body
state is denoted as a box with legs (double legs 
denote ket+bra),
Fig.~\ref{fig:tn}a, and an MPDO is expressed as a tensor network, where 
tensors are boxes, each leg denotes a tensor index, and connecting
legs corresponds to contraction, Fig.~\ref{fig:tn}b.

Let us briefly sketch the proof strategy: First, we show that for any
bipartition, a bound on the entanglement implies that the target state
$\rho$ can be well approximated by a low-rank decomposition across that
cut.  An area law thus implies that $\rho$ has low-rank approximations
across every cut. The crucial step will then be to merge these
approximations. To start, we will show how to merge two approximations in
such a way that (\emph{i}) we still obtain a good approximation and
(\emph{ii}) the internal structure of the two states is preserved 
(specifically, existing lower-rank approximations across other cuts),
as this allows us to iterate the procedure.  In a final step, we then show
how to nest this merging procedure in such a way as to obtain a good
MPDO approximation of the target state $\rho$.

We start by defining the $\alpha$-R\'enyi entanglement of purification
 $E_{p,\alpha}(\rho_{AB})$~\cite{terhal:entanglement-of-purification}.
 For a bipartite state $\rho_{AB}$, it is
given by
\begin{equation}
\label{eq:EoP-def}
E_{p,\alpha}(\rho_{AB}) = \min_{\ket{\psi}}
E_\alpha(\ket{\psi})\ ,
\end{equation}
where the minimum is taken over all purifications 
$\ket{\psi}_{AA'BB'}$ of $\rho_{AB}$, i.e.\
$\mathrm{tr}_{A'B'}\ket{\psi}\bra{\psi}=\rho_{AB}$, and
$E_\alpha(\ket\psi)=S_\alpha(\mathrm{tr}_{BB'}\ket{\psi}\bra{\psi})$, with
$S_{\alpha}(\rho)=\tfrac{1}{1-\alpha}\log(\mathrm{tr}\rho^\alpha)$, the
$\alpha$-R\'enyi entanglement entropy quantifying the pure state
entanglement between $AA'$ and $BB'$. For the remainder of this paper, we
restrict to $0\le\alpha<1$.

A key result from the pure state case~\cite{verstraete:faithfully} is that
a small $E_\alpha(\ket\psi)$ implies a rapid decay of the Schmidt
coefficients, and thus, there exists a low-rank
approximation to $\ket\psi$: Concretely, for any $D_p$ there
exists a $\ket{\chi_{D_p}}= \sum_{i=1}^{D_p} \ket{\chi^L_i}_{AA'}\ket{\chi^R_i}_{BB'}$
such that
\begin{equation}
\label{eq:D-scaling}
\eta 
:=1-|\langle\psi\vert\chi_{D_p}\rangle|^2
\le \left( 
    \frac{(1-\alpha)\,\exp[{E_{p,\alpha}(\rho_{AB})}]}{D_p}
\right)^{\frac{1-\alpha}{\alpha}}
\end{equation}
(and thus $D_p$ scales as an inverse polynomial in the error $\eta$).
This is equivalent to
\begin{equation}
\|\ket\psi\bra\psi - \ket{\chi_{D_p}}\bra{\chi_{D_p}}\|_1
\le 2\sqrt{1-
|\langle\psi\vert\chi_{D_p}\rangle|^2} = 2\sqrt{\eta}
\end{equation}
with $\|\cdot\|_1$ the trace norm (i.e.\ the sum of the singular
values)~\footnote{This is readily checked by working in the two-dimensional
space spanned by the two vectors; note that $\ket{\chi_D}$ need not be
normalized.}. By tracing $A'B'$, and using the fact
that tracing (as a completely positive trace preserving map) is contractive under the trace
norm, we arrive at
\begin{equation}
\label{eq:approx-error-onecut}
\delta:=\|\rho_{AB} - \sigma_D \|_1\le   2\sqrt{\eta}
\end{equation}
for some 
\begin{equation}
\label{eq:operator-schmidt-dec}
\sigma_{D} = \sum_{i,j=1}^{D_p} A_{ij}\otimes B_{ij}
\end{equation}
with rank $D=D_p^2$
(where $A_{ij}=\mathrm{tr}_{A'}\ket{\chi_i^L}\bra{\chi_j^L}$ and
$B_{ij}=\mathrm{tr}_{B'}\ket{\chi_i^R}\bra{\chi_j^R}$).

Let us now turn towards a spin chain of length $N$ whose state 
$\rho$ obeys an area law, that is, there is an $E_{\mathrm{max}}^\alpha$ such
that $E_{p,\alpha}(\rho_{AB})\le E^\alpha_\mathrm{max}$ for any bipartition
$A=1,\dots,L$, $B=L+1,\dots,N$. Combining
Eqs.~(\ref{eq:approx-error-onecut}) and (\ref{eq:D-scaling}), we
have that for each cut, there exists a rank $D=D_p^2$ decomposition
of the form 
\eqref{eq:operator-schmidt-dec} with trace norm error
\begin{equation}
\label{eq:delta-vs-D}
\delta \le 2\sqrt{\eta}
\le 2\left( 
    \frac{(1-\alpha)\,\exp[{E^\alpha_\mathrm{max}}]}{D^{1/2}}
\right)^{\frac{1-\alpha}{2\alpha}}\ .
\end{equation}
What remains to be seen is whether it is possible to merge these different low-rank
approximations. However, at this point 
we can no longer use the purifications to resort to the pure state result,
since the optimal purifications (minimizing $E_{p,\alpha}$) for
different cuts need not be related~\footnote{Different purifications
are related by a unitary on the purifying system $A'B'$, which however
mixes $A'$ and $B'$ and thus changes the entanglement properties.}. We
thus require a different approach.

\begin{figure}[t]
\includegraphics[width=\columnwidth]{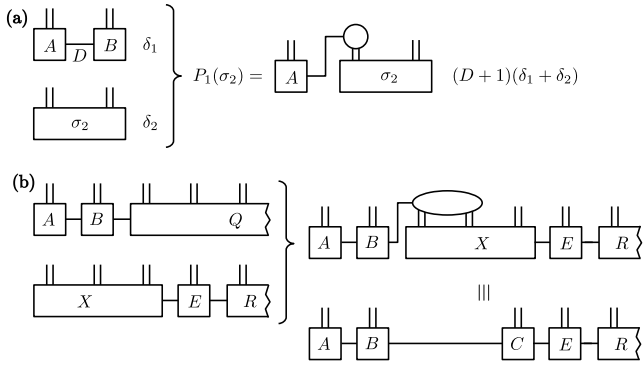}
\caption{\textbf{(a)} Merging two approximations with errors $\delta_1$
and $\delta_2$ yields an approximation with error
$(D+1)(\delta_1+\delta_2)$. Here, the circular tensor denotes $\sum_i
c_{ik}\hat A_i'$, cf.\ Eq.~\eqref{eq:P1-preserves-Ak}.  \textbf{(b)}~Merging preserves existing
cuts (i.e., local MPDO structure); here, we apply (a) to the first two
sites vs.\ the rest.}
\label{fig:merge}
\end{figure}

To start, consider a bipartite state $\rho\equiv\rho_{AB}$ (obtained
by blocking sites), a truncated
approximation \begin{equation}
\label{eq:sigma1}
\sigma_1 = \sum_{i=1}^{D} A_i\otimes B_i\ , \quad 
\|\rho-\sigma_1\|\le \delta_1\ ,
\end{equation}
and another approximation $\sigma_2$,
\begin{equation}
\|\rho-\sigma_2\|\le \delta_2\ ,
\end{equation}
obtained e.g.\ by truncating across a different cut. Let us now try to
connect those two approximations.  To this end, consider a (not necessarily
orthogonal) projection $P_1$ onto $\mathrm{span}\{A_i\}$,
$P_1(A_i)=A_i$, which can be written as $P_1(X) = \sum_{i=1}^D \hat
A_i\,\mathrm{tr}[(\hat A_i')^\dagger\,X]$
for some basis $\hat A_i=\sum c_{ik} A_k$ of $\mathrm{span}\{A_i\}$ and
some (dual) matrices $\hat A'_i$ satisfying $\mathrm{tr}[(\hat
A_i')^\dagger \hat A_j]=\delta_{ij}$.  $P_1$ can be naturally embedded into
the full space as
\begin{equation}
\label{eq:P1-fullspace}
P_1(X) = \sum_{i=1}^D \hat A_i\otimes 
\mathrm{tr}_A[(\hat A_i'\otimes\openone_B)^\dagger\,X]\ .
\end{equation}
Now consider
\begin{equation}
\label{eq:P1-preserves-Ak}
P_1(\sigma_2) = \sum_{i,k=1}^D c_{ik}A_k \otimes \mathrm{tr}_A[(\hat
A_i'\otimes\openone_B)^\dagger\sigma_2]\ ,
\end{equation}
see Fig.~\ref{fig:merge}a.
First, it also has rank $D$ across the cut; second, the left part is
spanned by $A_k$, and thus inherits the structure of the left part of
$\sigma_1$; and third, the right part is obtained from $\sigma_2$ by
tracing its left part with $(A_i')^\dagger$, and thus inherits the
structure of the right part of $\sigma_2$.  In
particular, if $\sigma_1$ and $\sigma_2$ have parts on the left and right,
respectively, which are already in Matrix Product form, both of these 
are inherited by $P_1(\sigma_2)$, see Fig.~\ref{fig:merge}b. We can then iterate
this scheme, starting from truncations at individual cuts, to obtain an
MPDO approximation.

What is the approximation error of the merged truncation $P_1(\sigma_2)$?
Using $P_1(\sigma_1)=\sigma_1$ from \eqref{eq:sigma1}, we have
\begin{align}
\label{eq:P1sigma2-rho}
\|P_1(\sigma_2)-\rho\|_1 & \le 
    \|P_1(\sigma_2)-P_1(\sigma_1)\|_1
	    +\|P_1(\sigma_1)-\rho\|_1
\quad
\nonumber
\\
& \le \|P_1(\sigma_2-\sigma_1)\|_1 + \|\sigma_1-\rho\|_1
\nonumber
\\
&\le \|P_1(X)\|_1 + \delta_1\ ,
\end{align}
with $X:=\sigma_2-\sigma_1$, $\|X\|_1\le
\delta_1+\delta_2$.
Starting from \eqref{eq:P1-fullspace},
a series of elementary inequalities~\footnote{%
Specifically,
\begin{align*}
\|P_1(X)\|_1 
    &\le \sum \|\hat A_i\otimes
\mathrm{tr}_A[(\hat A_i'\otimes\openone)^\dagger X]\|_1
\\
    &= \sum \|\hat A_i\|_1\,
\|\mathrm{tr}_A[(\hat A_i'\otimes\openone)^\dagger X]\|_1
\\
    &
\stackrel{\smash{(*)}}{\le}
\sum \|\hat A_i\|_1\,
\|(\hat A_i'\otimes\openone)^\dagger X]\|_1
\\
    &\le \sum \|\hat A_i\|_1\,
\|\hat A_i'\otimes\openone\|_\infty\, \|X\|_1
\\
    &\le \sum \|\hat A_i\|_1\,
\|\hat A_i'\|_\infty\, \|X\|_1\ ,
\end{align*}
where $(*)$ uses the contractivity of the partial trace.
}
gives 
\begin{equation}
\label{eq:P1-bound-sumD}
\|P_1(X)\|_1\le \sum_{i=1}^D \|\hat A_i\|_1\,
    \|\hat A_i'\|_\infty\, \|X\|_1\ .
\end{equation}
To keep $\|P_1(X)\|_1$ small, we thus ideally want to choose $\hat A_i$ and
$\hat A_i'$ such that $\|\hat A_i\|_1=\|\hat A_i'\|_\infty=1$ (this
is optimal as $\delta_{ij}=\mathrm{tr}[(\hat A_i')^\dagger \hat A_j]\le
\|\hat A_i\|_1\|\hat A_i'\|_\infty$).  It
turns out that such $\{\hat A_i\}$, $\{\hat A_i'\}$ indeed exist, a standard result in
functional analysis~\cite{werner:funktionalanalysis}: Choose a so-called Auerbach basis
of the normed space $\mathcal A = \mathrm{span}\{A_i\}$ with norm
$\|\cdot\|_1$, that is, a basis
$\{\hat A_i\}$ together with a set of linear functionals $\hat a_j':\mathcal
A\to\mathbb C$ such that $\hat a_j'(\hat A_i)=\delta_{ij}$ and $\|\hat
A_i\|=\|\hat a_j'\|=1$ (such a basis always exists)  and extend the
bounded functional $\hat a_j'$ to a bounded functional $\mathrm{tr}[(\hat
A_j')^\dagger\,\cdot\,]$, $\|\hat A_j'\|_\infty=1$,
whose existence is guaranteed by the Hahn-Banach theorem.

By inserting these $\{\hat A_i\}$, $\{\hat A_i'\}$ in Eq.~\eqref{eq:P1-bound-sumD},
we arrive at $\|P_1(X)\|_1\le D\,\|X\|_1$, which together with 
\eqref{eq:P1sigma2-rho} and $\|X\|_1\le\delta_1+\delta_2$ yields
\begin{equation}
\label{eq:error-merge2}
\|P_1(\sigma_2)-\rho\|_1 \le (D+1)\delta_1+D\delta_2 
\ .
\end{equation}
That is, we have merged the two approximations $\sigma_1$ and $\sigma_2$,
with new error as above; if both $\delta_1,\delta_2\le\delta$, the new
error is at most $(2D+1)\delta$.

\begin{figure}[b]
\includegraphics[width=\columnwidth]{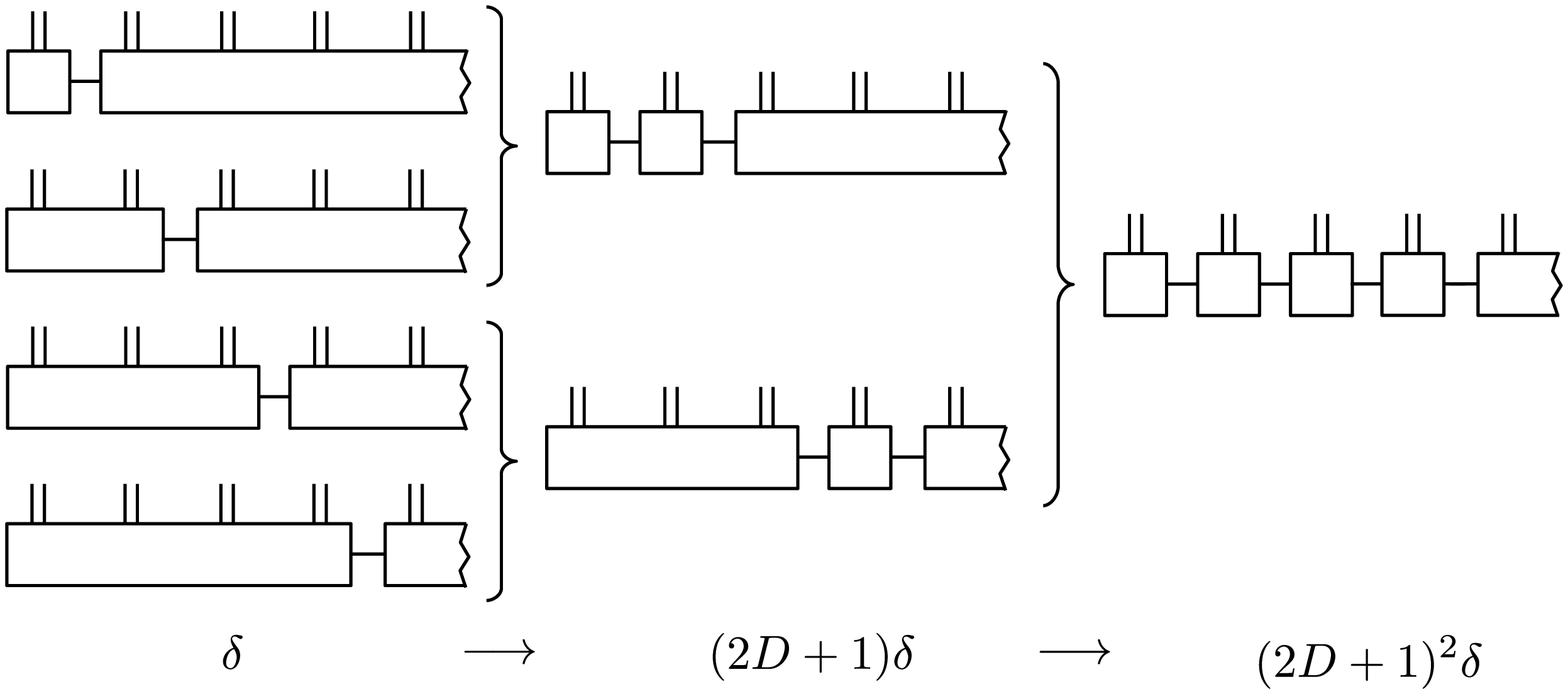}
\caption{Merging of cuts in a tree-like fashion. In each step,
the number of cuts is doubled, and the error grows by a factor of
$(2D+1)$. 
}
\label{fig:rg-merge}
\end{figure}

At this point, we can start concatenating truncations using
\eqref{eq:error-merge2}. However, we cannot do this sequentially as one
would do for the $2$-norm (where one can choose $P_1$ the orthogonal
projection for which $\|P_1(X)\|\le\|X\|$; note that this yields an
alternative proof for the result of Ref.~\onlinecite{verstraete:faithfully}):
The prefactor $(2D+1)$ would grow exponentially with the number of steps,
rendering the bound useless.
To overcome this issue, we choose a renormalization-like procedure, where
we concatenate the cuts in a tree-like fashion, as illustrated in
Fig.~\ref{fig:rg-merge}, using Eq.~\eqref{eq:error-merge2} in each step.  One can
readily check that each step doubles the number of cuts and multiplies the
error with $(2D+1)$; if the number of cuts is not a power of $2$, we can
start some branches of the tree later, or we can pad the spin chain with
trivial (uncorrelated) spins. For a chain of length $N$, this scheme thus
requires $K = \lceil\log_2 (N-1)\rceil\le (\log_2N) + 1$ steps, and thus
incurs a total error of 
\begin{equation}
\label{eq:epsilon-total-rg}
\varepsilon = (2D+1)^K \delta 
\le (2D+1)^{\log_2 N+1}\,\delta\ .
\end{equation}

We are now at the point where we can combine our results:
Combining Eqs.~\eqref{eq:delta-vs-D} and \eqref{eq:epsilon-total-rg}
yields
\begin{equation}
\varepsilon \le 2(2D+1)^{\log_2 N+1}
\left(\frac{(1-\alpha)e^{E^\alpha_\mathrm{max}}}{D^{1/2}}
\right)^{\frac{1-\alpha}{2\alpha}}\ .
\end{equation}
If we now -- following Eq.~\eqref{eq:eop-arealaw-bound}
 -- choose
\begin{equation}
\alpha=\frac{\lambda}{5\log_2 N}\,,\ 
e^{E_\mathrm{max}^\alpha}=N^c\,,\mbox{\ and\ }
D=N^\kappa\,,
\end{equation}
with $0<\lambda<1$, 
$\kappa>\tfrac{2c}{1-\lambda}$,
we have [using $1-\alpha\le1$,
$\tfrac{1-\alpha}{2\alpha}\ge 2\log_2 N/\lambda$, $2D+1<3D$, 
and $N^c/D^{1/2}\le1$]
\begin{align}
\varepsilon &\le 2(3D)^{\log_2 N+1} \left(\frac{N^c}{D^{1/2}}
\right)^{\frac{2\log_2 N}{\lambda}}
\\
& \le 6D\left(
    \frac{3 N^{2c/\lambda}}{D^{1/\lambda-1}}\right)^{\log_2N}
 \!\!\!\!= 6N^\kappa \left(\frac{3}{N^\Delta}\right)^{\log_2 N}
 \!\!\!\!\to 0
\end{align}
with $\Delta = \tfrac{1}{\lambda}(\kappa(1-\lambda)-2c)>0$, which thus
goes to zero super-polynomially as $N\to\infty$. This completes the proof
of our result.

In summary, in this work we have established when MPDOs can efficiently
describe quantum many-body systems.  We have derived conditions which a
state $\rho$ has to fulfill such that it can be approximated by an MPDO
with a polynomial bond dimension.  In particular, we have
shown that for a sequence of density
operators $\rho$ on a spin chain of length $N$, an entanglement area
law implies an efficient approximability of $\rho$ by 
MPDOs.  More concretely, we have considered a family of
area law bounds for the R\'enyi entanglement of purification which limit the
quantum correlations to grow at most logarithmically with the system size $N$; in
this setting, we have found that there exist MPDO approximations to $\rho$
with a bond dimension which grows polynomially in the system size $N$, and
for which  the approximation error decreases faster than any inverse
polynomial in $N$. This shows that MPDOs provide a faithful approximation
to density operators which satisfy an area law, and are thus well suited
for the numerical simulation as well as analytical study of such systems.

\emph{Acknowledgements.---}This work has received support from the
European Union's Horizon 2020 program through the ERC StG WASCOSYS
(No.~636201), the ERC CoG GAPS (No.~648913), and the ERC AdG QENOCOBA
(No.~742102), from the DFG (German Research Foundation) under Germany’s
Excellence Strategy (EXC-2111 -- 390814868), and through the Severo Ochoa
project SEV-2015-0554 (MINECO).

\bibliography{../../bibtex/all}

\end{document}